\documentclass[aps,showpacs,prb,reprint,longbibliography,footinbib,citeautoscript,raggedbottom]{revtex4-2}
\usepackage{graphicx}
\usepackage{dcolumn}
\usepackage{bm}
\usepackage{microtype,lmodern}
\usepackage{xr}
\usepackage[x11names,dvipsnames]{xcolor}
\usepackage[bookmarks=false,colorlinks]{hyperref}
\hypersetup{linkcolor=magenta,citecolor=MidnightBlue,filecolor=Plum,urlcolor=MidnightBlue}

\begin{document}

\newcommand{\jmr}[1]{\textcolor{blue}{#1}}
\renewcommand\sectionautorefname{Section}
\renewcommand\subsectionautorefname{Section}

\preprint{APS/123-QED}

\title{Can Autonomous LLM Agents Execute Multireference Quantum Chemistry Calculations?}

\author{Victor Chang Lee}
\affiliation{Department of Materials Science and Engineering, Northwestern University, Evanston, IL, USA}

\author{James M.\ Rondinelli}
\email{jrondinelli@northwestern.edu}
\affiliation{Department of Materials Science and Engineering, Northwestern University, Evanston, IL, USA}
\date{\today}

\begin{abstract}

Multireference electronic-structure calculations remain difficult to automate because critical workflow decisions, including active-space selection, state averaging, convergence recovery, and state identification, traditionally rely on expert judgment. Here, we investigate whether an autonomous large language model (LLM) agent can perform these tasks without human intervention. The agent selects active spaces using literature-grounded analogies or explicitly documented chemical reasoning, generates and submits ORCA calculations, analyzes outputs, and records all decisions in an auditable reasoning log. Benchmarking against 558 vertical transition energies (VTEs) from QUESTDB shows that an unguided baseline agent achieves 24.9\% coverage with a mean absolute error (MAE) of 0.373 eV. Introducing a structured decision ladder increases coverage to 44.1\% while reducing the MAE to 0.339 eV. The largest gains are observed for double and Rydberg excitations, demonstrating that expert-informed procedural guidance substantially improves active-space construction and state identification. When provided with complete workflow information, the agent successfully reproduces published QUEST calculations with an MAE of only 23 meV, resolving 75\% of target configurations within seven attempts. These results demonstrate that contemporary LLM agents can autonomously execute and reproduce complex multireference quantum-chemical workflows, while highlighting the importance of structured reasoning frameworks for achieving reliable high-throughput and high-fidelity electronic-structure calculations.

\end{abstract}

\maketitle

\section{Introduction}

In recent years, the generation of large-scale computational databases has emerged as a major breakthrough in materials research and molecular chemistry. At the first-principles level, density functional theory (DFT) serves as the primary method for data generation \cite{burke2012,Hohenberg1964,Kohn1996}. Benefiting from standardized computational pipelines that are straightforward to automate, along with a reasonable computational cost, DFT has enabled high-throughput databases that collectively span millions of systems, such as Meta FAIR’s OMol25 \cite{levine2026}, the Materials Project \cite{Horton2025}, and the Open Quantum Materials Database (OQMD) \cite{Kirklin2015}, among others. 
However, standard exchange-correlation functionals to DFT  fail systematically when static electron correlation is strong, impacting predictive capability of excited states \cite{Saade2024}, $d$- and $f$-block complexes \cite{Khurana2026,Morrillo2025}, diradicals, bioinorganic active sites \cite{Barreiro2024}, and photochemical reaction paths \cite{Miao2024}. 

Multiconfigurational (multireference) electronic structure methods, such as the Complete Active Space Self-Consistent Field (CASSCF) \cite{Siegbahn1981,Dang2021}, more accurately  describe these strongly correlated regimes. The so-called active space in CASSCF is a subset of electrons and orbitals that are optimized using a full configuration iteration (FCI) to obtain the configuration iteration (CI) coefficients and the multiconfigurational wave function. To account for remaining dynamic correlation, CASSCF calculations are refined using multireference perturbative methods. One of the most prominent approaches is the $N$-electron valence state perturbation theory (NEVPT2).

A primary bottleneck in applying multiconfigurational methods at scale lies in determining which electrons and orbitals need to be treated explicitly within the correlated wavefunction \cite{wardzala2026}. Omitting essential orbitals fails to capture critical static correlation, whereas an excessively large active space rapidly becomes computationally intractable. Traditionally, active space selection relies on manual choices grounded in chemical intuition and literature precedents; it is also often biased towards a particular property of interest, presenting a major barrier to the large-scale generation of high-accuracy data for multiple or yet specified purposes.

To minimize reliance on user intervention, several automated selection protocols have been introduced \cite{wardzala2026}. The two prominent examples are the  atomic valence active space (AVAS) and AutoCAS \cite{avas2017, avas2019, autocas2016,autocas2019}. The former uses a set of atomic orbitals (AO) of a minimal basis set that are sufficient to qualitatively represent the final CASSCF active orbitals while the later uses large-active-space density matrix renormalization group (DMRG) \cite{Chan2004} calculations to identify orbitals characterized by strong correlation. Other efforts have tailored active space automation to high-throughput workflows and machine learning (ML) applications. To that end, Gagliardi and coworkers developed an automated protocol for molecular dynamics and ML potentials that generates initial active space guesses by interpolating molecular orbital coefficient matrices from previously optimized wavefunctions \cite{jeong2020}. More recently, the same group developed an automated algorithm for high-throughput magnetic property calculations of lanthanides complexes \cite{zhou2026}.

Recent advances in large language models (LLMs) offer a promising pathway to overcome this bottleneck. LLMs have demonstrated rapid improvements in technical reading comprehension, structured information extraction, and scientific reasoning \cite{Gottweis2026, Ghareeb2026,Shi2025}. %
In computational chemistry, LLMs have been studied as tools for molecular structure generation and prediction \cite{Cavanagh2026,cavanagh2026-2}.
More recently, LLM agents have recently been suggested for multireference computational chemistry tasks \cite{chen2026}. However, the full extent of their operational autonomy and decision making fidelity has not been systematically evaluated. Moving beyond static script generation or simple code execution requires quantifying how effectively an agent can independently handle non-trivial chemical judgments, including active-space selection, state-averaging schemes, and dynamic failure recovery, especially when confronted with challenging electronic structures. Leveraging these capabilities, we benchmark an autonomous LLM agent on executing complex multiconfigurational CASSCF calculations. By combining literature-derived domain context with dynamic decision scaffolding, the agent autonomously selects active spaces, constructs and executes a quantum chemistry software package. This framework allows us to systematically evaluate the agent's ability to exercise expert-level chemical judgment, establishing both its baseline predictive accuracy and the key challenges in scaling autonomous multiconfigurational workflows.

\section{Agent Architecture and Benchmark Design}
\label{agent}

\subsection{Overview}

\begin{figure}
    \centering
    \includegraphics[width=1\linewidth]{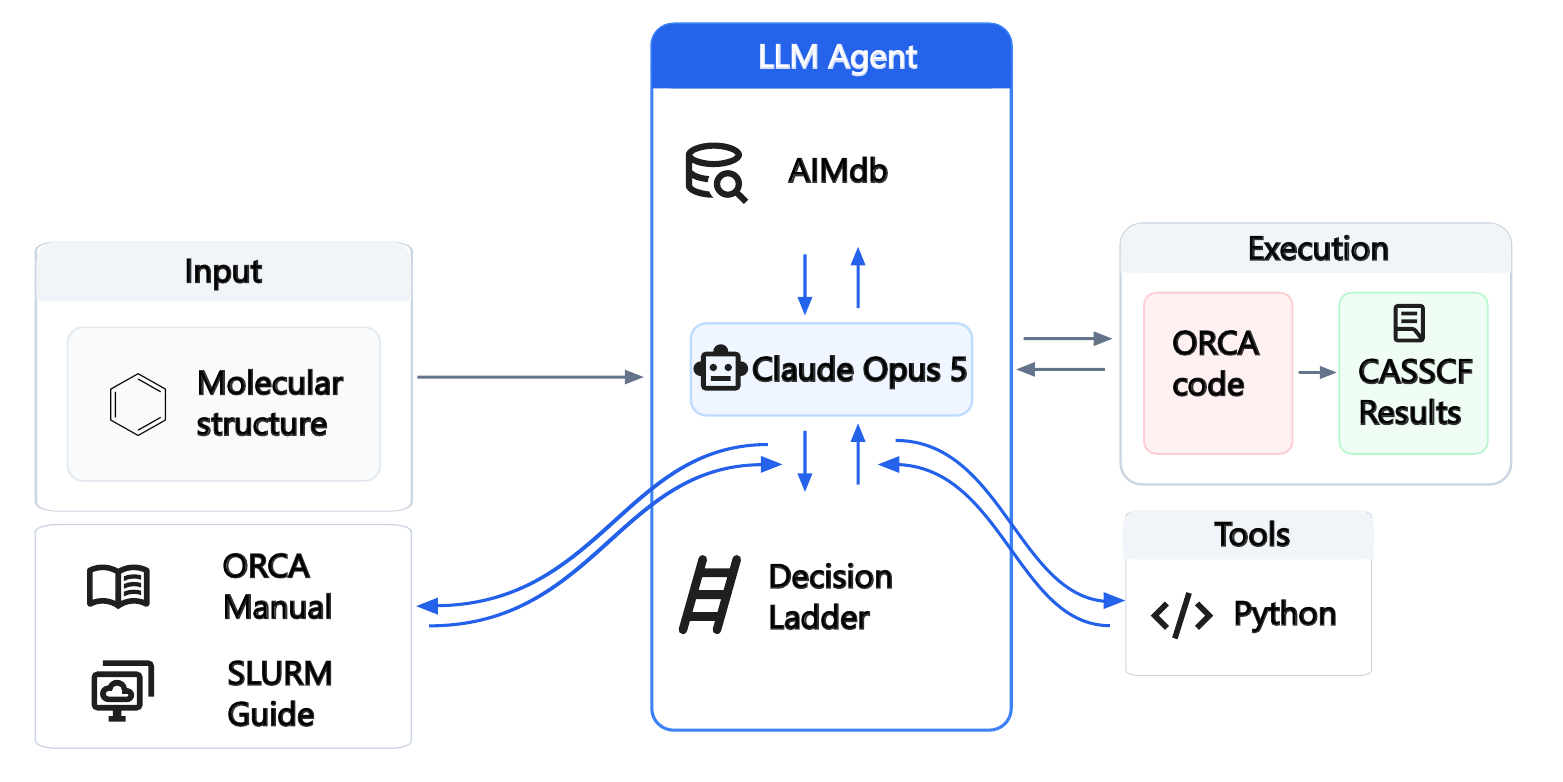}
    \caption{Overview of the autonomous LLM agent workflow for quantum chemical calculations. Input molecular structures, along with reference documentation (ORCA manual and SLURM workload manager guide), are passed to the core agent powered by Claude Opus 5. The agent interacts with an internal database (AIMdb) and operates under a structured decision ladder to guide execution logic. During execution, the agent leverages Python scripting tools, consults its knowledge sources to generate CASSCF results.}
    \label{fig:agent}
\end{figure}

As illustrated in \autoref{fig:agent}, the autonomous LLM decision agent operates as a transparent reasoning engine rather than a black-box automation pipeline. The agent evaluates each molecule's geometry, consults a curated literature database, constructs and submits ORCA v6.1 \cite{ORCA6} inputs. 
In CASSCF calculations, a geometry file alone is insufficient to start a calculation; it requires, at minimum, the electron/orbital counts of the active space, the multiplicities to compute, and the number of roots per multiplicity. These initial selections are informed by the knowledge sources provided to the model (see below). The LLM agent dynamically constructs the Python tools it deems necessary to complete its assigned task.
To that end, the LLM agent makes all substantive chemistry decisions and records its reasoning in an append-only, per-molecule audit log. The escalation logic (decision ladder) is kept strictly advisory and is never automatically executed. This design choice, enforced in code, stems from the realization that per-molecule quantum chemical judgment requires dynamic reasoning that static programs cannot provide.

\subsection{Knowledge Sources}

Benchmarking LLM-guided workflows is challenging because undisclosed training data can introduce  data-contamination effects \cite{zhou2023}. 
To mitigate this risk, we supply the agent with a curated reference database and explicitly instruct  it to exclude any prior knowledge or literature precedents contained in QUESTDB \cite{Loos2025}. The literature database was extracted from AIMdb \cite{aimdb2026}, and is searched using the compound name or formula, metal center, $d/f$ electron count, and ligand environment when a transition-metal or lanthanide/actinide center is present. For each matching entry, the agent retrieves the active-space size, orbital description, computational protocol, multiplicities, method, and associated citation.

Analogical reasoning is defined narrowly and explicitly in the agent policy. For a transition-metal and $f$-electron complex, a literature precedent is considered analogous only when it shares the same metal, oxidation state, and coordination environment.
For organic and main-group systems, analogy requires the same functional group and conjugation pattern. Every active-space recommendation derived from an analogous precedent  must cite the corresponding database entry. When no genuinely analogous examples are found, the agent policy defaults to sizing the active space directly from general quantum-chemical principles. The resulting record is explicitly tagged so that readers can distinguish literature-based from agent-derived  choices throughout the workflow and in all downstream data exports.

When the target active orbitals can be specified in terms of atomic orbitals (AOs), the policy requires building the active space using the Atomic Valence Active Space (AVAS) approach rather than specifying  explicitly the  electron/orbital counts ($n_\mathrm{el}$, $n_\mathrm{orb}$) and allowing ORCA to populate the window strictly from the orbital-energy frontier \cite{avas2017,avas2019}. Because AVAS derives the active space from chemically motivated AO targets, agreement with cited active-space sizes provides confidence in the selected space, while disagreements prompt additional examination of the orbital content and space definition.

\subsection{Governance and Audit Trail}
\label{sec:governance}

All chemical and technical decisions are delegated to the LLM agent without requiring a human approval. However, every decision must be grounded in literature precedent, the escalation ladder, or explicitly-flagged chemical intuition, and logged with reasoning to a per-molecule, in
append-only files.

The agent is required to stop and flag a particular calculation for four
categories only: a genuine give-up condition (the attempt budget
exhausted with every distinct action tried and recorded), 
literature conflict too large to explain (\textbf{Rule F}, see below), anything resembling data
corruption and infrastructure failure outside the ladder's scope. No
early stopping condition exists for a repeated diagnosis and
a recurring failure mode is explicitly defined as evidence that the previous fix was wrong, not that the molecule is intractable. The decision ladder enumerates the number of genuinely distinct actions available under a single diagnosis to clarify that a diagnosis is rarely exhausted after two or three attempts.

Overruling the ladder or a cited literature entry is permitted but carries three obligations: 

\begin{enumerate}
    \item State explicitly what is being overruled and provide the rational for doing so.
    \item Ground the override in something that can be checked against the same output (e.g., composition, occupation, symmetry, etc.).
    \item Document the override appropriately rather than framing it as established precedent.
\end{enumerate}

\subsection{Convergence criteria}
\label{sec:readbacks}

To determined whether a CASSCF calculation has both numerically converged and produced a physically meaningful wave function, the agent is required to evaluate four independent validation criteria:

\begin{enumerate}
  \item \textbf{Total energy relative to Hartree-Fock (HF):} The CASSCF energy is compared against the corresponding HF energy obtained with the same molecular geometry and basis set. For a given state, the CASSCF energy should generally be lower than the HF reference. Although state-averaged calculations spanning many roots may occasionally yield energies slightly above the HF value, substantial increases indicate an incorrectly defined active space, convergence failure, or wave-function instability.
  \item \textbf{Ground-state configuration-interaction (CI)  weight:}
  The weight of the dominant electronic configuration is monitored as a measure of static correlation. A leading configuration with a CI weight approaching unity indicates that the active space is contributing little to the wave function and that the calculation may be effectively single reference. Rather than applying a fixed universal threshold, the policy evaluates the ground state dominant CI weights relative to normally correlating attempts for that molecule.
  \item \textbf{Irreducible representation (irrep) distribution:} The symmetry distribution of active orbitals is compared with the intended active-space. Agreement in the total number of orbitals is insufficient; the orbitals must also occupy the correct irreps and correspond to the intended chemical subspace. An apparently correct irrep can still arise from an active space containing orbitals of incorrect chemical character.
  \item \textbf{Orbital indices and CI-configuration positions:}
    Molecular orbital (MO) indices and  CI-configuration identifies are valid only for the exact wave-function state that generated it. Because orbital ordering and CI string positions can change between optimization stages, state averages, and restarted calculations, the policy requires continuous verification of orbital indices and CI string positions at every decision point. Historical index lists are never assumed to remain valid, as doing so can lead to orbital rotations, active-space modifications, or state assignments being applied to unintended targets.
\end{enumerate}

\subsection{Decision ladder}
\label{sec:ladder}

The decision ladder defines a prioritized hierarchy of corrective and diagnostic actions used by the agent when a calculation deviates from the desired outcome. The rules are evaluated sequentially from highest to lowest priority, ensuring that fundamental execution and numerical issues are resolved before chemically motivated refinements are considered. Rather than serving as rigid, hard-coded constraints, the ladder functions as a collection of expert-informed heuristics and policy recommendations that guide autonomous decision-making while preserving flexibility across diverse chemical systems.

\begin{itemize} 
    \item \textbf{Rule A: Job execution failure.} Detect and resolve scheduler- or infrastructure-level failures, including SLURM submission errors, node failures, wall-time limits, and missing output files. 
    \item \textbf{Rule B: SCF non-convergence.} Address failures of the underlying Hartree--Fock procedure through appropriate convergence aids, restart strategies, or modifications to the initial guess. 
    \item \textbf{Rule C: CASSCF non-convergence.} Remedy failures of the multiconfigurational optimization, including problematic orbital rotations, inadequate active spaces, or unstable state averaging. 
    \item \textbf{Rule D: Convergence to an incorrect electronic state.} Identify situations in which the calculation converges formally but yields a chemically incorrect solution, such as an unintended occupation pattern, spin state, or orbital character. 
    \item \textbf{Rule E: Incomplete correlation treatment.} Recognize cases where convergence is achieved but the selected active space or state-averaging window is insufficient to capture the relevant electronic structure. 
    \item \textbf{Rule F: Strong disagreement with literature precedent.} Compare converged results with analogous literature benchmarks and investigate significant discrepancies in active-space composition, state ordering, or qualitative electronic structure. 
    \item \textbf{Rule G: Rydberg or diffuse-state contamination.} Detect and appropriately treat diffuse excited states that may require specialized basis functions, enlarged active spaces, or revised state-selection strategies. 
    \item \textbf{Rule H: Successful completion.} Confirm that the calculation has converged, the active orbitals possess the intended chemical character, orbital occupations are physically reasonable, and the results are consistent with available literature expectations. 
\end{itemize}

The agent's ability to make chemically and  physically informed calculations is concentrated within \textbf{Rules} \textbf{D}, \textbf{E}, and \textbf{G}, which address scenarios where formal numerical convergence alone is insufficient to guarantee a physically meaningful wave function. These rules, described below, govern incorrect electronic states, incomplete correlation treatment, and diffuse-state character.

\subsubsection*{\textbf{Rule D}: Convergence to an Incorrect Electronic State}
\label{ss:rule-d}

During orbital optimization, the solver can gradually rotate  a physically relevant active orbital into the inactive or virtual space while simultaneously replacing it with an unintended orbital. Consequently, the calculation may formally converge, yet the final active space can differ substantially from the intended one and contain orbitals with incorrect chemical character. To detect such failures, \textbf{Rule D} performs a comprehensive orbital-composition analysis based on three complementary diagonistics: natural orbital occupation numbers (NOON), L\"{o}wdin orbital decomposition analysis, and  geometric atomic-contribution analysis. Together, these metrics assess both the correlation character and chemical identity of the active orbitals.

When an incorrect active-space composition is identified, the primary corrective action is orbital rotation. Orbital rotations are used to bias the calculations toward a target electronic state by exchanging pairs of MOs from a previous  calculation. Within the decision ladder, orbital rotation is treated as a core state-targeting mechanism rather than a post hoc workaround. To maintain physical consistency, however,  rotations are strictly constrained from converged orbitals. Furthermore, when point-group symmetry is enabled, rotations must strictly preserve spatial symmetry by operating within individual irreps. To further prevent active-space drift, this rule pairs orbital rotations are coupled with level shifting and adaptive solver-selection strategies that stabilize retention of the intended active-space manifold.

\begin{figure*}
    \centering
    \includegraphics[width=0.78\linewidth]{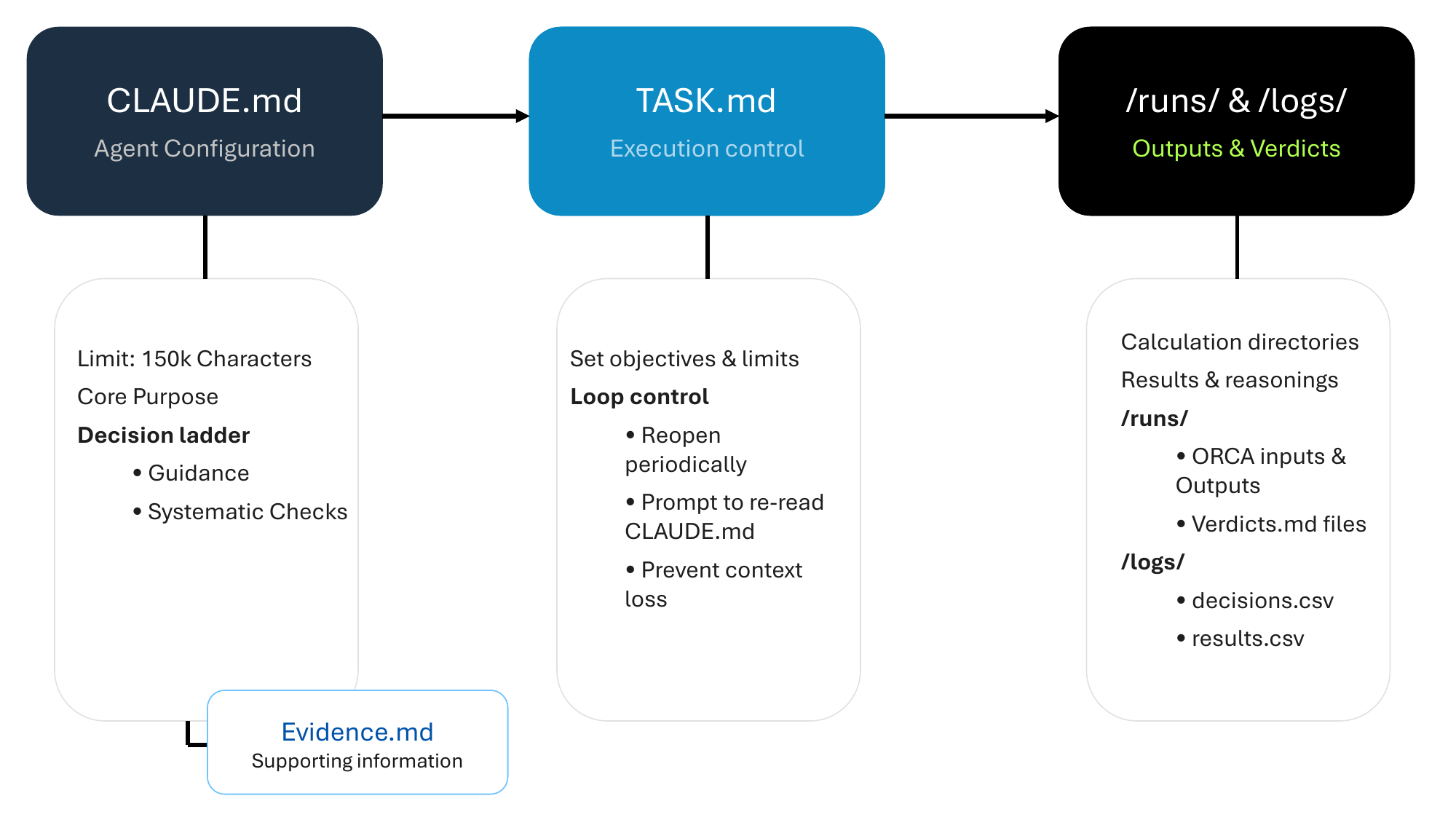}\vspace{-8pt}
    \caption{Agent architecture overview showing interactions between agent configuration (\texttt{CLAUDE.md} and \texttt{Evidence.md}), execution loop control (\texttt{TASK.md}), and output directories containing calculation runs (\texttt{/runs/}) and structured verdict logs (\texttt{/logs/}).}
    \label{fig:claude}
\end{figure*}

\subsubsection*{\textbf{Rule E}: Clean convergence with an Incomplete Correlation Space or Truncated Energy-root Window}
\label{ss:rule-e}

A CASSCF calculation may converge smoothly with a modest number of active electrons, orbitals, and state-averaged roots yet still fail to provide an adequate description of the electronic structure. Such cases arise when the active space is too small to capture the relevant correlation effects or when the state-averaging window excludes energetically accessible states that influence the target solution.
\textbf{Rule E} therefore probes the completeness of both the active space and the root window. For active-space validation, the agent examines the natural orbital occupation numbers (NOONs), paying particular attention to the least occupied active orbital. 
A small but non-negligible occupation suggests that additional correlating orbitals may be contributing to the wave function. In response, the subsequent calculation expands the active space and records the occupations of the newly introduced orbitals. The decision to retain the larger active space is based on the observed occupation pattern, e.g., mutual degeneracy, rather than a fixed threshold.

Selection of the state-averaging window is governed by the energetic and symmetry relationships among the target states rather than by an arbitrary root count. When near-degenerate states are present, the window is expanded to ensure a balanced description of the relevant electronic manifold. Further expansion is constrained by available literature precedent and chemical justification. Beyond this regime, the policy favors performing separate calculations for distinct sets of states instead of indefinitely widening a single state-averaged calculation. Excessively large state-averaging windows can degrade the quality of the optimized molecular orbitals, as the orbitals must simultaneously accommodate an increasingly diverse set of electronic states. Consistent with this expectation, calculations performed with narrow state-averaging windows typically yield slightly lower energies for a given target state than otherwise identical calculations employing much broader root windows.

\subsubsection*{\textbf{Rule G}: Rydberg (diffuse) states}
\label{ss:rule-g}

The preceding rules are designed to identify deficiencies in active-space composition, orbital occupations, and state selection; however, they are not sufficient for reliably detecting Rydberg excitations. Unlike valence excitations, Rydberg states are distinguished primarily by the large spatial extent and diffuse character of the orbitals involved rather than by their occupation patterns alone. Consequently, identifying such states requires explicit analysis of orbital diffuseness in addition to conventional active-space diagnostics.

To address this limitation, the policy employs ORCA's multiconfigurational random phase approximation (MCRPA) linear-response module using a converged CASSCF reference wave function. MCRPA is used exclusively as a diagnostic tool to identify candidate excited states and the orbitals that contribute to them. Because MCRPA excitation energies are obtained within a CASSCF linear-response framework, they do not recover the dynamic correlation subsequently captured by NEVPT2 and are therefore not used for reporting final excitation energies.

Once candidate orbitals have been identified, their diffuse character is quantified by evaluating the smallest Gaussian exponent associated with the relevant atomic center and angular-momentum channel. Orbitals with significant contributions from highly diffuse basis functions are classified as Rydberg candidates and ranked accordingly. The identified diffuse orbitals are then incorporated into an expanded active space, after which the target state is recomputed using SA-CASSCF followed by NEVPT2. This procedure enables the agent to describe Rydberg states within a physically meaningful active space while retaining the quantitative accuracy of the final correlated treatment.

\begin{figure*}
    \centering
    \includegraphics[width=0.72\linewidth]{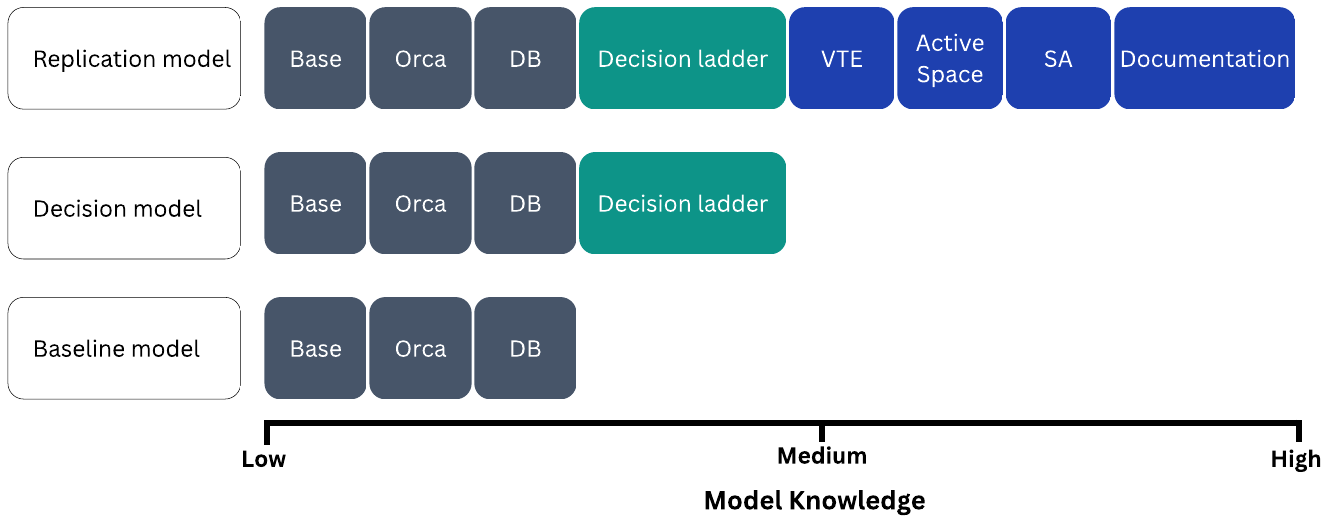}
    \caption{Framework for evaluating LLM agents performance across varying levels of domain context. The benchmark progresses along a gradient of model knowledge from minimal contextual information to extensive domain-specific guidance, enabling systematic assessment of replication fidelity and agent reliability. (DB: Database; VTE: Vertical transition energies; SA: State Average)}
    \label{fig:test}
\end{figure*}

\subsection{Agent Implementation} 

The Claude Opus 5 model was deployed using a modular instruction architecture (\autoref{fig:claude}). 
Claude Code initializes each agent session from a \texttt{CLAUDE.md} file, which defines the system instructions and operational framework governing the agent's behavior. In practice, however, the size of this file is constrained, with performance often degrading as the instruction set grows beyond approximately 150,000 characters. Furthermore, during extended execution, newly generated content progressively occupies the context window (capped at 1 million tokens), increasing the risk that critical instructions, policy details, or reasoning guidance will receive less attention or be displaced from the model's active context.

To improve instruction persistence and reduce context-management failures, the agent architecture separates information across three specialized files:
\begin{itemize}
    \item \texttt{CLAUDE.md}: Defines the system context, operational boundaries, and core policies (including the decision ladder and ORCA's manual checks).
    \item \texttt{TASK.md}: Manages execution state and objective tracking, forcing the agent to re-read system instructions at each iteration loop.
    \item \texttt{EVIDENCE.md}: Serves as an auxiliary knowledge store, holding detailed reasoning, supporting context, and rule-specific justifications for the primary policies in \texttt{CLAUDE.md}.
\end{itemize}
This seperation minimizes competition for context-window resources, improves retention of critical operational policies, and provides a scalable framework for long-running autonomous calculations.

\section{Computational Methods}

Complete active space self-consistent field (CASSCF) calculations followed by strongly contracted N-electron valence state perturbation theory (SC-NEVPT2) were performed using the ORCA v6.1 software package \cite{ORCA,ORCA6,Angeli2001}. For all calculations, the aug-cc-pVTZ basis set was employed, consistent with the default protocol of QUESTDB. To accelerate the calculation of two-electron integrals, the resolution-of-identity chain-of-spheres exchange (RIJCOSX) approximation  was utilized \cite{Kossmann2010} together with auxiliary basis sets automatically generated via ORCA’s AutoAux feature. Active space construction and orbital selection were governed autonomously by an LLM agent workflow powered by Anthropic's Claude Opus 5 model \cite{anthropic2026claude}.

To evaluate the autonomous decision-making performance of the agent, a curated benchmark set of 84 molecules was selected from the 2025 QUESTDB\cite{Loos2025} obtained with similar CASSCF/SC-NEVPT2 protocol. Although the reference benchmark calculations in QUESTDB were generated using MolPro \cite{molpro}, all CASSCF/SC-NEVPT2 calculations in this work were executed using the ORCA v6.1 software package to align with our local computational workflow.

\section{Results and Discussion}

\subsection{Benchmark Framework}

To evaluate the ability of the LLM agent to autonomously perform multireference electronic-structure calculations, we benchmarked the workflow against the vertical transition energy (VTE) dataset contained in  QUESTDB using only the molecular structure files. Three information regimes were considered (\autoref{fig:test})
\begin{enumerate}
\item \textbf{Baseline model}: Access to the ORCA documentation and the literature database only.
\item \textbf{Decision model}: The Baseline model supplemented with the decision ladder described in \autoref{sec:ladder}.
\item \textbf{Replication model}: Access to all available QUESTDB knowledge, including all information required to reproduce published calculations.
\end{enumerate}
These three tiers define a controlled knowledge gradient, ranging from general quantum-chemistry resources to complete task-specific information. This framework allows the effects of domain context on agent performance to be isolated and quantified.

The \textbf{Baseline model} establishes the capabilities of an essential unguided LLM agent operating with access only to standard software documentation and literature precedents. 
The ORCA manual provides methodological and implementation details, while the literature database supplies externally grounded examples of active spaces, state selections, and computational protocols. This design also serves as a partial control for potential training-data contamination, allowing decisions supported by explicit literature retrieval to be distinguished from those derived from the model's internal chemical reasoning \cite{aimdb2026}.
Across the benchmark set, the agent identified direct literature analogues for the vast majority of molecules.
Only four molecules from the QUESTDB benchmark dataset were deemed dissimilar from available precedents that active-space selection and state-averaging choices were were based primarily on chemical intuition rather that literature-grounded analogy \cite{Hapka2019Second,Hoffman1999X,Kerridge2013Oxidation,Lennartz2013Computational,Li2025Benchmark,Feixas2011Electron,Lykhin2021Dipole,Monte2023Quantification,Romeu2025Electronic,Sepali2024Fully,Sousa1997Theoretical,AN2025Opacities,Anglada1996Unimolecular,Anglada2010Dissociation,Arenas2002Carbene,Babb2017Radiative,Battaglia2021Role,Bauschlicher1987Theoretical,Bera2008Born,Berger1998Calculation,Besley1999Ab,Bhattacharjee2019Resolving,Blaise2024Isomerization,Bone2025Benchmarking,Buma1995Vibronic,Buzsaki2026Trajectory,Calio2022Nonadiabatic,Chachisvilis1999Femtosecond,Cui2011Channels,Ding2003Combined,Fang2010Wavelength,Feher1995Ab,Freitas2023Espectroscopia,Ganyushin2013Fully,Ghosh2008Orbital,Gonzalez2001Vtst,GonzalezVazquez2008Casscf,Greiner2024Mbe,Guan2014Combined,Guareschi2013Ground,Guareschi2014Solvent,Gudem2025Mechanism,Guo2025Approximation,Hashimoto1996Theoretical,HelmichParis2019Benchmarks,HelmichParis2025Excited,HelmichParis2025Two,Hohenstein2015Analytic,Hu2018Inclusion,Jangrouei2022Dispersion,Jiang2024Quantum,Jovanovic2017Performance,Keller2015Selection,Koga1991Comparison,Kubelka2001Ab,Kuhlman2012Between,Lan2014Toward,Langhoff1987Theoretical,Langhoff1988Theoretical,Lavorato1996Observation,Luo2012Ground,Lykhin2021Dipoleb,Mai2019Influence,Maksic2000Correlation,Malcolm1997Combining,Meng2011Casscf,Menger2023Fluorescence,Nakajima2001Theoretical,Nishimoto2025Analytic,Oliva2005Low,Pelaez2007Dependence,Picconi2018Photodissociation,Ponzi2016Photoionization,Potts2001Improved,Rowell2020Structural,Saade2024Excited,Saha2026Direct,Santolini2015Photochemical,Sayos2001Lowest,Scott2020Analytic,SerranoAndres2002Electronic,Olsen2014Canonical,Sharma2019Density,Shiozaki2013Pyrazine,Shiozaki2016Hyperfine,Siegbahn1981Complete,Sobolewski1995Ab,Soto2021Sa,Stock1995Resonance,Sugisaki2009Ab,Varras2018Explanation,Ventura2024New,Venturini2002Caspt,Wei2007Theoretical,Xu2022Multireference,Zammit2022Comprehensive,Zhang2003Ab,Zhang2006Electronic}.

The \textbf{Decision model} augments the Baseline model with a decision ladder, providing a structured set of expert-informed policies for diagnosing and correcting common failures in mulitreference calculation workflows. Importantly, the decision ladder neither overrides the LLM's intrinsic reasoning process nor prescribes specific calculation inputs; it also does not have final authority over task-level decisions. 
Instead, it supplies procedural guidance while leaving final decisions to the agent, thereby isolating the value of formalized domain expertise independent of detailed task-specific instructions.

Finally, the \textbf{Replication model} provides the agent with all workflow information to reproduce the QUESTDB results, including access to their repository resources, calculation protocols, and supporting and documentation. This regime represents an upper bound on the contextual information available to the agent and evaluates its ability to reproduce published multireference calculations when complete domain knowledge is accessible.

\subsection{Impact of the Decision Ladder}

To quantify the role of structural domain guidance, we compare the performance of the  Baseline and Decision models for predicting VTEs from molecular structures using CASSCF/SC-NEVPT2.
Statistical error metrics, including the mean absolute error (MAE) and coefficient of determination ($R^2$), were computed exclusively for VTEs whose identity against the reference could be unambiguously verified against the reference benchmark. VTEs identified solely by energy ordering or those belonging to a manifold of degenerate/near-degenerate states with ambiguous assignments were  excluded from the error analysis and classified as unidentified. Consequently, failures in state assignment are reflected through the reduced coverage rather than artificially skewing the reported MAE or $R^2$ values.

\begin{figure}
    \centering
    \includegraphics[width=1\linewidth]{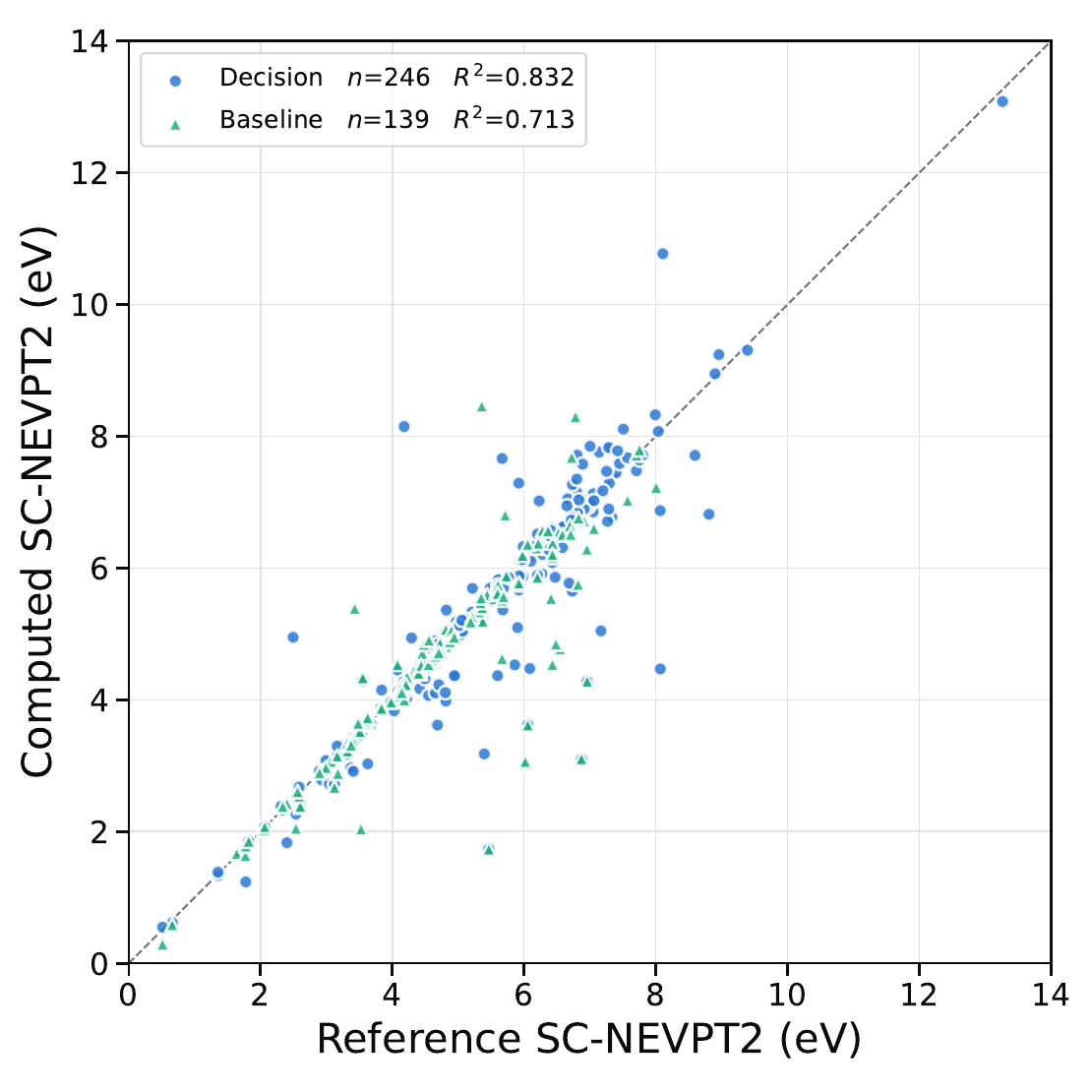}
    \caption{Comparison of computed and reference vertical transition energies for the Baseline and Decision models.
    The dashed line denotes perfect agreement.}
    \label{fig:scatter}
\end{figure}

The addition of the decision ladder significantly improves the overall performance. 
As shown in \autoref{fig:scatter}, $R^2=0.713$ for the Baseline model increases to $R^2=0.832$ for the Decision model, while the number of verified VTEs $N_\mathrm{scored}$ increased from  139 to 246.
This simultaneous improvement in accuracy and coverage indicates that the procedural  knowledge encoded within the decision ladder enables the agents to solve and execute a substantially larger fraction of multireference quantum chemistry problems without sacrificing predictive quality. This is further supported by the distribution of errors. In the Baseline model, 22 out of 139 scored VTEs (15.8\%) exhibited absolute errors greater than 0.5 eV, with a maximum error of 3.97 eV. In comparison, the Decision model increased the number of successfully identified VTEs to 246, of which 46 (18.7\%) exhibited absolute errors exceeding 0.5 eV and a maximum error of 3.76 eV. Although the decision ladder greatly expands the range of molecules that can be successfully treated, these results indicate that additional refinement of the decision policies are necessary to systematically reduce the largest energy errors.

\begin{table*}
  \centering
  \caption{Evaluation metrics by VTE excitation classes for the Baseline and Decision models. $N_{\text{scored}}$ denotes the number of verified computed vertical transitions included in the statistical analysis, and $N_{\text{refs}}$ is the total number of reference transitions in QUESTDB.}
  \label{tab:metrics_mae_comparison}
  \begin{ruledtabular}
  \begin{tabular}{l c  ccc  ccc}
    & & \multicolumn{3}{c}{\textbf{Baseline}} & \multicolumn{3}{c}{\textbf{Decision}} \\
    \cline{3-5}\cline{6-8}
    Class & $N_{\text{refs}}$ & $N_{\text{scored}}$ & Coverage (\%)& MAE (eV) & $N_{\text{scored}}$ & Coverage (\%) & MAE (eV) \\
    \hline
    Single  & 356 & 124 & 34.8 & 0.384 & 165 & 46.3 & 0.332 \\
    Double  & 30  & 0   &  0.0 & --    & 14  & 46.7 & 0.488 \\
    Rydberg & 124 & 6   &  4.8 & 0.116 & 57  & 46.0 & 0.308 \\
    TM      & 48  & 9   & 18.8 & 0.393 & 10  & 20.8 & 0.429 \\
    \hline
    Total   & 558 & 139 & 24.9 & 0.373 & 246 & 44.1 & 0.339 \\
  \end{tabular}
  \end{ruledtabular}
\end{table*}

Performance is evaluated separately for each class of VTEs so that accuracy and coverage are assessed independently (\autoref{tab:metrics_mae_comparison}). Across the full benchmark out of $N_{\text{refs}}=558$ reference VTEs, the Decision model increased dataset coverage from $24.9\%$ ($N_{\text{scored}}=138$ in the Baseline model) to $44.1\%$ while simultaneously reducing the overall MAE of $0.373~\text{eV}$ to $0.339~\text{eV}$. However, performance varied substantially depending on the underlying electronic character of the VTE. For successfully scored VTEs, the agent was reasonable stable and there were no  catastrophic outliers (\autoref{fig:scatter}).

The largest volume of scored cases belonged to single (valence) excitations ($N_{\text{refs}} = 352$). The Decision model increased both metrics substantially, yielding $165$ scored VTEs ($46.3\%$ coverage) and reducing the MAE to $0.332~\text{eV}$ compared to the Baseline model with $124$ scored VTEs ($34.8\%$ coverage) and  an MAE of $0.384~\text{eV}$.
Double excitations proved more difficult for the Baseline model, which failed to produce any verified assignments. In contrast, the Decision model successfully identified $14$ out of the $30$ double excitations ($46.7\%$ coverage). This result highlights the importance of explicit policies for diagnosing incomplete active spaces and incorrect state assignments.

\subsection{Recovery of Rydberg States}
The most dramatic improvement was observed for Rydberg excitations, where coverage increased from only $4.8\%$ to $46.0\%$ while keeping a MAE of $0.308~\text{eV}$ comparable to that of single excitations. 
This behavior is consistent with the design of \textbf{Rule G}, which explicitly introduces orbital-diffuseness analysis and targeted active-space expansion for Rydberg-state identification. Without these diagnostics, the Baseline model rarely selected the diffuse orbitals required to describe Rydberg excitations. Transition-metal systems remained the most challenging category, exhibiting only modest improvements in coverage and similar MAEs between the two models.
These results demonstrate that identifying diffuse orbital character is a critical capability that is not naturally captured through occupation analysis alone.

\subsection{Transition-Metal Systems}
Transition-metal excitations remained the most challenging class in the benchmark. Although the decision-guided workflow substantially improved the accuracy of successfully identified states, increasing reliability came at the expense of benchmark coverage. For the subset of four VTEs successfully evaluated by both workflows, the Decision model consistently outperformed the Baseline model, reducing the MAE from 0.348 to 0.217~eV.

The reduction in coverage highlights a  difficulty of transition-metal electronic structures. Unlike the predominantly organic systems found elsewhere in the benchmark, transition-metal can exhibit near-degenerate $d$-orbital manifolds, competing spin states, and complex excited manifolds and state-averaging. These features increase ambiguity in active-space selection and state assignment, making it difficult to distinguish physically meaningful solutions from formally converged but chemically incorrect wave functions. The improved accuracy achieved by the Decision model suggests that the additional validation procedures successfully filtered out a fraction of these incorrect solutions. However, the accompanying reduction in coverage indicates that the current decision policies remain insufficiently robust for general transition-metal chemistry. Future improvements will likely require more specialized active-space construction strategies, stronger state-tracking procedures, and a larger body of transition-metal-specific literature precedents.

\subsection{Coverage}

To assess the robustness of our workflow, coverage was evaluated across all 84 target molecules. The Baseline model failed to reproduce a single reference-matched VTE for 33 molecules, whereas the Decision model failed on 29 molecules. Of these failures, 23 were shared between both models, while 7 and 3 were unique to the Baseline and Decision models, respectively.
The decision ladder substantially improved molecular-level coverage. The Decision model achieved an MAE below 0.3 eV for 38 molecules, including 15 for which all reference VTEs were successfully reproduced. In comparison, the Baseline model achieved an MAE below 0.3 eV for 35 molecules, but only 3 reproduced the complete reference VTE set. The increased breadth of the Decision model is exemplified by triazine, for which all 14 reference VTEs were recovered, whereas the best Baseline result reproduced only 6 of 8 VTEs for acrolein.

Across molecules with at least one successfully identified VTE, the Decision model reproduced 71.7\% of reference VTEs per molecule compared with 46.7\% for the Baseline model. For the subset of 109 VTEs identified by both workflows, the two models exhibited comparable accuracy, with MAEs of 0.268 and 0.242~eV for the Decision and Baseline models, respectively. The largest error in both workflows occurred for cyclopentadienone, which exhibited MAEs of 3.763 (Decision model) and 3.774~eV (Baseline model).

The increase in coverage was accompanied by a higher computational cost. The Baseline model completed 131 calculations spanning 84 distinct structure, active-space, and state-averaging configurations. In contrast, the Decision model executed 576 calculations across 148 configurations and reached the maximum eight-attempt limit for two molecules without obtaining a successful result. Nevertheless, both workflows remained conservative relative to the QUESTDB benchmark, which contains 502 distinct active-space and state-averaging protocols.

\subsection{Failure Modes and Limitations}

The dominant failure mode for both models was the inability to 
identify an active space and state-averaging strategy
to reproduce the reference electronic states. These failures led into incorrect underlying orbital manifolds, missed VTEs, and incomplete descriptions of electronically complex systems, particularly molecules containing transition-metals.

More broadly, most unsuccessful benchmark cases arose from incorrect or ambiguous state assignments rather than large numerical errors in successfully identified states. This observation suggests that future improvements are more likely to come from enhanced orbital interpretation, state tracking, and active-space construction than from incremental refinements in the underlying electronic-structure methods. Accurately identifying the desired electronic states remains a difficult problem even for expert practitioners, especially when little a priori information is available.

Despite these limitations, the benchmark demonstrates that LLM agents can achieve reasonable quantitative accuracy when appropriate active spaces and state assignments are obtained. More importantly, the results show that structured decision policies substantially improve the autonomous execution of multireference quantum-chemical workflows by increasing both the reliability and breadth of successful calculations. This highlight opportunities for improving state-selection strategies rather than fundamental limitations of the agent-based framework itself.

\subsection{Replication Test}

\begin{figure}
    \centering
    \includegraphics[width=1\linewidth]{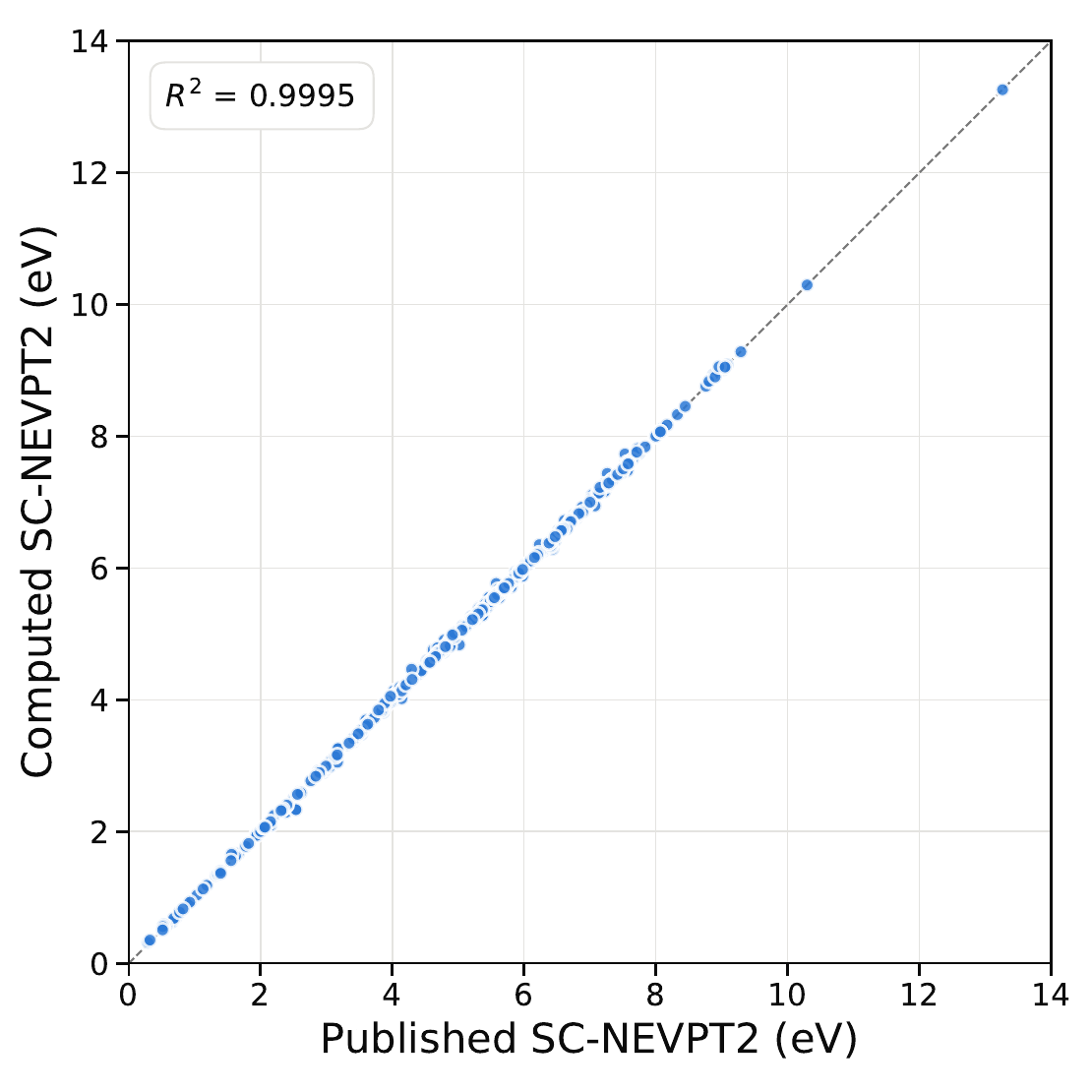}
    \caption{Comparison of computed and reference vertical transition energies for the Replication model. The dashed line denotes perfect agreement.}
    \label{fig:replication}
\end{figure}

To establish an upper bound on agent performance, we evaluated a Replication model provided with all information necessary to reproduce the QUESTDB reference calculations. This included the active space definitions,  state-averaging specifications by irrep, and the associated publications and supplementary information.
As shown in \autoref{fig:replication}, the replicated calculations exhibit excellent agreement with the reference data, without evidence of systematic bias or significant outliers. This result shows that the LLM agent can reliably reproduce published multireference workflows when provided with the necessary computational context.

\begin{table}
\centering
  \caption{Evaluation metrics by VTE excitation classes for the Replication model. $N_{\text{scored}}$ denotes  the number of verified computed vertical transitions, and $N_{\text{refs}}$ the total number of reference transitions in QUESTDB.}
  \label{tab:replication_mae}
  \begin{ruledtabular}
  \begin{tabular}{l c  ccc }
    & & \multicolumn{3}{c}{\textbf{Replication}}  \\
    \cline{3-5}
    Class & $N_{\text{refs}}$ & $N_{\text{scored}}$ & Coverage (\%)& MAE (eV) \\
    \hline
    Single  & 356 & 266 & 74.7 & 0.023  \\
    Double  & 30  & 19   &  63.3 & 0.019     \\
    Rydberg & 124 & 61   &  49.2 & 0.021\\
    TM      & 48  & 32   & 66.7 & 0.037  \\
    \hline
    Total   & 558 & 378 & 67.7 & 0.023  \\
  \end{tabular}
  \end{ruledtabular}
\end{table}

Unlike the previous benchmarks, which only verified VTE identity matches, successful replication additionally required reproducing the published active-space and state-averaging protocols. 
Under these stricter criteria, the Replication model achieved an overall MAE of 23 meV, representing an order-of-magnitude improvement over both the Baseline and Decision models. Performance was remarkably consistent across excitation classes, with transition-metal systems showing only a modest increase in error (\autoref{tab:replication_mae}).

In total, 201 VTEs were reproduced with errors below 5 meV, including 51 reproduced exactly. The remaining overall MAE of 0.023 eV is primarily due to differences in state-average weighting between ORCA and MOLPRO. 56 states were computing using weighting schemes that differed from the original MOLPRO calculations, resulting in a higher average MAE of 45.1 meV. Whereas MOLPRO applies equal weighting to all roots, ORCA weights roots by blocks. Although this discrepancy is documented in the decision ladder, it was not strictly enforced within the decision-making process, leading to it being overlooked in multiple instances. In contrast, when the agent explicitly corrected the state-average weights to match the reference protocol (22 states), the average MAE decreased to 8.1~meV. Thus, the residual discrepancies arise primarily from workflow implementation details rather than failures in state identification or active-space selection.

Overall, coverage reached 67.7\%,  with most unreproduced VTEs attributable to the imposed limit of eight calculation attempts: 117 VTEs reached the attempt cap without a successful result, 51 produced incorrect solutions, 9 converged without yielding a usable assignment, and only 3 failed due to other errors. Thus, the principal constraint on coverage was not accuracy but the computational budget allocated to autonomous exploration.

While increasing the allowed attempts could improve coverage, the resulting computational cost must be factored into evaluations of LLM replication capabilities as the computational cost of replication was substantial. The agent submitted 2,459 ORCA jobs, of which 1,253 successfully reproduced the target VTEs. For these successful runs, the agent needed 3.48 calculation attempts on average to successfully reproduce a calculation. The largest contributor to unsuccessful calculations was the eight-attempt cap, accounting for 776 jobs. Nevertheless, 42\% of target specifications ($n=210$) were reproduced on the first attempt, typically when MP2 natural orbitals already generated the correct active space. Moreover, 75\% of successful reproductions were completed within seven attempts, and 77\% within the imposed limit of eight attempts.
These results indicate that roughly one-third of benchmark calculations required at least one corrective intervention by the agent.

\begin{figure}
    \centering
    \includegraphics[width=1\linewidth]{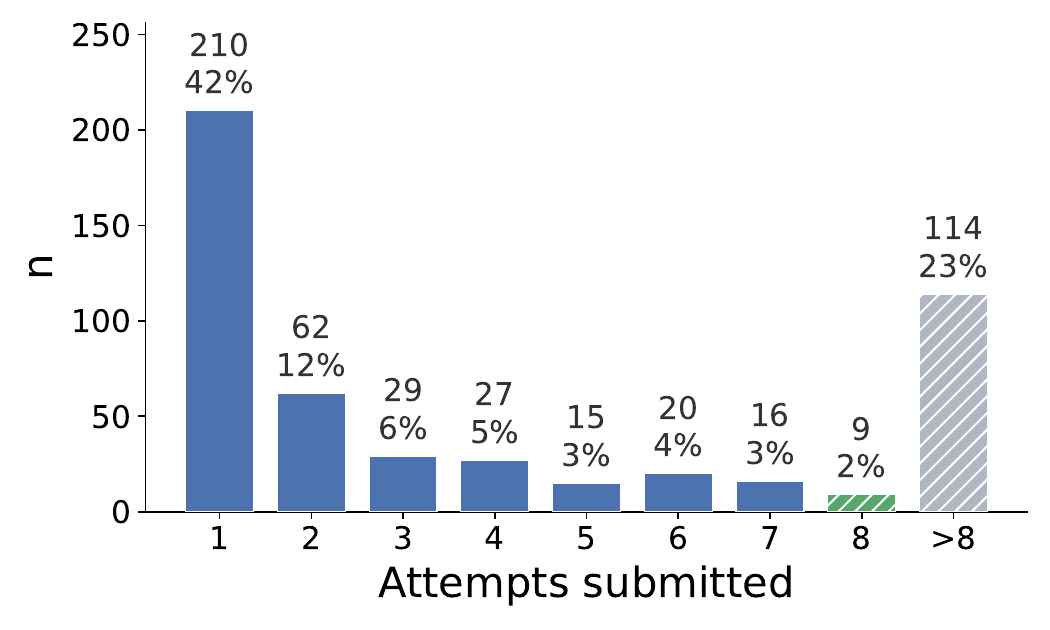}\vspace{-6pt}
    \caption{Distribution of calculation attempts executed by the LLM agent per target specification. Percentages and absolute counts ($n$) are shown above each bar. The imposed attempt threshold is capped at 8 ($n = 9$, 2\%), while the gray-hatched bar ($>8$, $n = 114$, 23\%) corresponds to specifications that failed to converge within the maximum allowed attempt limit and would require additional calculations to complete.}
    \label{fig:attempts}
\end{figure}

The distribution in \autoref{fig:attempts} also suggests diminishing returns beyond approximately eight attempts, as increasingly large computational effort is required to recover a progressively smaller fraction of remaining targets. Consequently, future improvements are likely to be achieved more effectively through refinement of the decision policies than by simply increasing the allowed number of attempts.
Additionally, the agent could be used as a screening or ranking tool to filter out routine calculations while reserving challenging cases for human experts.

\section{Conclusions}
We benchmarked an autonomous LLM agent capable of executing multireference quantum-chemical workflows, including active-space selection, ORCA input generation, output analysis, convergence recovery, and electronic-state identification. Across 558 reference vertical transition energies from QUESTDB, introducing a structured decision ladder increased benchmark coverage from 24.9\% to 44.1\% while reducing the overall MAE from 0.373 to 0.339~eV. The largest gains were observed for double and Rydberg excitations, demonstrating that expert-informed decision policies can substantially improve the reliability and breadth of autonomous multireference calculations. Transition-metal systems remained the most challenging class, highlighting the need for more specialized active-space construction, state-tracking procedures, and domain-specific guidance. 

The replication benchmark establishes a practical upper bound on agent performance when complete contextual information is available. Under these conditions, the agent reproduced published calculations with an overall MAE of only 23~meV, with 42\% of target calculations completed on the first attempt and 75\% within seven attempts. 
Our results show that LLM agents are capable of reliably reproducing complex multireference workflows. The dominant remaining limitations arise from workflow efficiency and search strategy rather than an inability to recover the correct electronic structure. The ability to autonomously replicate published calculations creates new opportunities for automated verification, cross-platform benchmarking, and large-scale reproducibility studies of quantum chemical data. 

To that end, our results demonstrate that the effectiveness of LLM agents in scientific computing depends strongly on the quality of the reasoning framework provided to them. General-purpose language models possess substantial chemical knowledge but often lack the structured decision-making processes required for demanding multistep workflows. Embedding expert knowledge in the form of decision ladders, validation criteria, and domain-specific retrieval systems offers a practical route toward improving reliability while retaining the flexibility of foundation models. 
Future progress will likely come from agent architectures that combine specialized scientific knowledge bases, more sophisticated state identification and active-space selection strategies, and tighter integration between machine reasoning and electronic-structure theory. Such developments could substantially reduce computational overhead while increasing coverage and accuracy, enabling autonomous, high-throughput multireference quantum chemistry and accelerating the generation, verification, and dissemination of computational benchmark data.

\section*{Code and data availability}

All data presented in this work were generated using publicly available software. The LLM agent and documentation, the generated data and LLM reasoning is available at [Placeholder].

\section*{Acknowledgments}

This material is based upon work supported by the U.S. Department of Energy, Office of Science, Office of Basic Energy Sciences under Award Number DE-SC0025176. This research was supported in part through the computational resources and staff contributions provided for the Quest high performance computing facility at Northwestern University which is jointly supported by the Office of the Provost, the Office for Research, and Northwestern University Information Technology. This work used Anvil at Purdue University through allocation PHY250069 from the Advanced Cyberinfrastructure Coordination Ecosystem: Services \& Support (ACCESS) program, which is supported by U.S. National Science Foundation grants \#2138259, \#2138286, \#2138307, \#2137603, and \#2138296. During the preparation of this manuscript/study, the author(s) used Microsoft Copilot (institutional instance) for the purposes of editing and language polishing. No new content was generated, and the authors have reviewed and edited the output and take full responsibility for the content of this publication.

\section*{Author Contributions}

V.C.L.\ - Writing - Original Draft, Conceptualization, Methodology, Software, Formal analysis, Investigation, Visualization, Resources. 
J.M.R.\ -  Writing - Original Draft , Supervision, Project administration, Funding acquisition.

\section*{Competing interest}
The authors declare no competing interests.


\bibliography{aipsamp_pdflatex}

\end{document}